\documentclass[11pt]{article}
\usepackage{array}
\usepackage{amsmath}
\usepackage{amssymb}	
\newcommand{\vev}[1]{\left\langle #1 \right\rangle}

\begin{document}
\begin{titlepage}
\begin{center}
\hfill arXiv:\,0707.0429\\ 
\vskip 1cm
\begin{LARGE}
\textbf{The Axion From Five-Dimensional Supergravity} 
\end{LARGE}\\
\vspace{1.0cm}
Sean McReynolds \footnote{sean.mcreynolds@mib.infn.it}\\
\vspace{.35cm}
\emph{University of Milano-Bicocca and INFN Milano-Bicocca\\
Piazza della Scienza 3, 20126 Milano, Italy}\\
\vspace{1.0cm} {\bf Abstract}
\end{center}
We consider the axion arising from five-dimensional supergravity in the presence of boundaries.  We find the approximate bosonic effective action to estimate the lower bound on the axion coupling scale $M_{PQ}$ with a flat bulk.  With a warped bulk, one can obtain an $M_{PQ}$ within the standard window; this puts a bound on the required curvature scale relative to the proper separation between boundaries.  We comment on the scalar potential that may ruin the strong-CP resolution, and the effective derivative coupling to matter in 5D hypermultiplets.

\vfill {\flushleft {July 2007}}
\end{titlepage}

\section{Introduction}\label{sec:intro}

Field theories on a spacetime manifold with boundaries, modeled by the topological space $\mathbb{R}^{D}\times S^{1}/\mathbb{Z}_{2}$, have in the recent past been seriously considered in phenomenological model building.  A supersymmetric extension of these models is often still considered for stabilizing the new hierarchy problems that arise, including the radion field whose vacuum expectation value is the distance between boundaries~\cite{radion, susyRS}.  Ultimately, such models should be embedded in supergravity. Supergravity on such spacetimes is also of interest in string/M-theoretic compactifications.  In this paper, we'll consider the axion that naturally arises in five-dimensional Yang-Mills-Einstein supergravity theories (YMESGTs) on a manifold with boundaries, modeled by $\mathbb{R}^{4}\times S^{1}/\mathbb{Z}_{2}$; on these boundaries a restricted set of gauge fields survive.    

Due to QCD instantons, the effective Lagrangian of the Standard Model contains a CP-violating term
\begin{equation}
\mathcal{L}_{F\tilde{F}}=- \frac{1}{64\pi^{2}g^{2}}\theta \epsilon^{\mu\nu\rho\sigma}\mbox{Tr}[\mathcal{F}_{\mu\nu}\mathcal{F}_{\rho\sigma}],\label{theta}
\end{equation}
where $\theta$ parametrizes the instanton vacuum, the trace is in the adjoint representation of $SU(3)\times SU(2)\times U(1)$ and $g$ is the 4D gauge coupling.  From~\cite{Harris:1999jx}, the bound on the parameter can be estimated to be $\theta\leq \mathcal{O}(10^{-10})$ and the lack of explanation for such a small dimensionless number is known as the strong-CP problem. 

One possible resolution comes in the form of an axion field $\mathcal{A}$ with the coupling 
\begin{equation}
\mathcal{L}_{\mathcal{A}}= -\frac{1}{64\pi^{2}g^{2}}\frac{\mathcal{A}}{M_{PQ}} \epsilon^{\mu\nu\rho\sigma}\mbox{Tr}[\mathcal{F}_{\mu\nu}\mathcal{F}_{\rho\sigma}], \label{axion}
\end{equation}
where $M_{PQ}$ is a characteristic energy scale.  The low energy theory then contains the effective vacuum parameter $\theta+\vev{\mathcal{A}}/M_{PQ}$, which vanishes due to an instanton-induced scalar potential for $\mathcal{A}$.  In the original idea put forth by Peccei and Quinn~\cite{Peccei:1977ur}, Weinberg~\cite{Weinberg:1977ma}, and Wilczek~\cite{Wilczek:1977pj}, the axion is the pseudo-Goldstone boson of an approximate, rigid $U(1)_{PQ}$ ``Peccei-Quinn" symmetry that is broken at a scale $M_{PQ}$. Astrophysical considerations provide a lower bound $M_{PQ}\hspace{-1mm}\geq \hspace{-1mm}\mathcal{O}(10^{10})\mbox{GeV}$ \cite{Kim:1986ax}, while cosmological arguments provide an upper bound $M_{PQ}\hspace{-1mm}\leq \hspace{-1mm}\mathcal{O}(10^{12})\mbox{GeV}$~\cite{astrophysical}.  In the original axion scenario, the Peccei-Quinn symmetry breaking scale can't be characterized by the breakdown of chiral symmetry and the appearance of the instanton vacuum.  Whatever the origin of the axion is, the new physics must generate an intermediate energy scale. 

The axionic coupling in Eq.~(\ref{axion}) can arise from higher-dimensional Chern-Simons type couplings, where the axions are vector fields with components in the extra dimensions. Chern-Simons type couplings, in turn, can arise upon integrating out massive fermions in odd dimensions; 5D examples of axions arising in such a way can be found in~\cite{warpedaxion}.  However, such couplings are present in classical supergravity (and string) theories, which is the framework that we are using for higher dimensional scenarios.  

A generic drawback in these situations is that the Chern-Simons coupling comes with the gravitational scale so that $M_{PQ}$ is typically too large. One can look for ways of lowering $M_{PQ}$ to within the standard window; a warped bulk is useful in this regard~\cite{warpedaxion}.  Alternatively, one can look for ways of raising the upper bound allowed for $M_{PQ}$~\cite{Banks:2003sx}. In some supersymmetric models~\cite{Banks:2002sd}, a scale as large as $10^{16}\mbox{GeV}$ may be allowed.  Also, the standard cosmological arguments assume that $\vev{\mathcal{A}}/M_{PQ}\sim 1$ in the early universe~\cite{astrophysical}; if it had $\vev{\mathcal{A}}/M_{PQ}\sim 10^{-3}-10^{-2}$, and the mass of the axion is larger than the Hubble parameter, then $M_{PQ}\leq \mathcal{O}(10^{16}-10^{18})\mbox{GeV}$ is allowed.  Furthermore, the evolution in the energy density of the 4D axion field involves dynamics in the full 5D theory. 

In Sect.~\ref{sec:framework} we discuss how the approximate 4D effective theory arises from a 5D Yang-Mills-Einstein supergravity theory.  In Sect.~\ref{sec:qcdaxion} we obtain estimates of the axion coupling scale $M_{PQ}$ in a flat and warped bulk.  This follows the idea of~\cite{warpedaxion} in the warped case, though the details differ. In Sect.~\ref{sec:scalarpot} we discuss the presence of the potential that arises from gauge couplings of 5D charged scalars, and in which the scalars composing the axion generally appear.  Finally, in Sect.~\ref{sec:matter} we check that the derivative coupling to matter in bulk hypermultiplets allows the standard low energy analysis of axion couplings.                  

\section{Framework}\label{sec:framework}

\subsection{5D $\mathcal{N}=2$ vector-coupled supergravity}\label{sec:sugra}

Five-dimensional $\mathcal{N}=2$ Maxwell-Einstein supergravity theories (MESGTs)~\cite{Gunaydin:1983bi} consist of the $\mathcal{N}=2$ bare gravity supermultiplet $\{e^{\mu}_{m},\Psi^{\iota}_{\mu},A^{0}_{\mu}\}$ coupled to $n_{V}$ $\mathcal{N}=2$ vector multiplets $\{A^{i'}_{\mu},\lambda^{\iota\,p},\phi^{x}\}$, where $\hat{\mu}=1,\ldots,5$ is a curved spacetime index; $m=\bar{1},\ldots,\bar{5}$ is a flat spacetime index; $x=1,\ldots,n_{V}$ is a curved index of the real target space $\mathcal{M}_{R}$; $p=1,\ldots,n_{V}$ is a flat index for the target space; $i'=1,\ldots,n_{V}$; and $\iota=1,2$ is a doublet index for the $SU(2)_{R}$ automorphism group of the supersymmetry algebra.  The $n_{V}$ scalars sit in $n_{V}+1$ functions $h^{I'}$, which are determined by a cubic polynomial $\mathcal{V}=C_{I'J'K'}h^{I'}h^{J'}h^{K'}=1$, where $C_{I'J'K'}$ is a rank-3 symmetric tensor that completely determines the MESGT.  In ``canonical" form, its components are  
\begin{equation}
C_{000}=1\;\;\;\;C_{00i'}=0\;\;\;\;C_{0i'j'}=-\frac{1}{2}\delta_{i'j'}\;\;\;\;C_{i'j'k'}=\mbox{arbitrary},
\label{canonical}\end{equation}
so that the $C_{i'j'k'}$ contain the choice of MESGT. The isometry group $Iso(\mathcal{M}_{R})$ of $\mathcal{M}_{R}$ contains the (possibly trivial) invariance group $G$ of $C_{I'J'K'}$, which is a rigid symmetry group of the MESGT action on $\mathbb{R}^{5}$.  A subgroup $K\subset G\times SU(2)_{R}$ can be gauged if the $n_{V}+1$ vector fields form a representation containing the adjoint of $K$.  Theories resulting from gaugings in $G$ are called Yang-Mills-Einstein (YMESGTs), while gaugings in $SU(2)_{R}$ are called ``gauged supergravities"~\cite{Gunaydin:1984ak}.  In general, then, theories with $K\subset G\times SU(2)_{R}$ are called ``gauged YMESGTs".      
 
Let $I=(0,i)$, where $i$ without a prime denotes K-non-singlets (other than the adjoint), and K-singlets (other than the graviphoton if it is one).  The bosonic Lagrangian for a 5D $\mathcal{N}=2$ YMESGT is~\cite{Gunaydin:1984ak} 
\begin{equation}\begin{split}
\hat{e}^{-1}\mathcal{L}_{5}=&-\frac{1}{2\hat{\kappa}^{2}}\hat{\mathcal{R}}-\frac{1}{4\hat{g}^{2}}\stackrel{\circ}{a}_{IJ}\mathcal{F}^{I}_{\hat{\mu}\hat{\nu}}\mathcal{F}^{J}_{\hat{\rho}\hat{\sigma}}\,\hat{g}^{\hat{\mu}\hat{\rho}}\hat{g}^{\hat{\rho}\hat{\sigma}} - \frac{3}{4\hat{\kappa}^{2}}\stackrel{\circ}{a}_{IJ}D_{\hat{\mu}}h^{I}D_{\hat{\nu}}h^{J}\,\hat{g}^{\hat{\mu}\hat{\nu}}\\
&+\frac{\hat{\kappa}\hat{e}^{-1}}{6\sqrt{6}\,\hat{g}^{3}}C_{IJK}\epsilon^{\hat{\mu}\hat{\nu}\hat{\rho}\hat{\sigma}\hat{\lambda}}\{F^{I}_{\hat{\mu}\hat{\nu}}F^{J}_{\hat{\rho}\hat{\sigma}}A^{K}_{\hat{\lambda}}+\cdots\}
\end{split}\end{equation} 
where 
\[
\mathcal{F}^{I}_{\hat{\mu}\hat{\nu}}=(\partial_{\hat{\mu}}A^{I}_{\hat{\nu}}-\partial_{\hat{\nu}}A^{I}_{\hat{\mu}})+A^{J}_{\hat{\mu}}\,f^{I}_{JK}A^{K}_{\hat{\nu}}\;\;\;\;\;\mbox{and}\;\;\;\;\;
D_{\hat{\mu}}h^{I}=\partial_{\hat{\mu}}h^{I}+A^{J}_{\hat{\mu}}\,f^{I}_{JK}h^{K},
\]           
hats indicate 5D objects, and the ellipsis indicates the additional terms in the non-abelian ``Chern-Simons" term.  The Riemannian metric in the kinetic terms is $\stackrel{\circ}{a}_{IJ}=-\frac{1}{3}\partial_{I}\partial_{J}\ln\mathcal{V}|_{\mathcal{V}=1}$.  

\subsection{YMESGT on $\mathbb{R}^{4}\times S^{1}/\mathbb{Z}_{2}$}\label{sec:orbifold}

We will consider this theory on $\mathbb{R}^{4}\times S^{1}/\mathbb{Z}_{2}$, where $S^{1}$ is coordinatized as $x^{5}\in [-\pi R, \pi R]$ and $\mathbb{Z}_{2}$ acts as $x^{5}\mapsto -x^{5}$ with fixed points at $\{0\},\{\pi R\}$.  Working on the covering space $\mathbb{R}^{4}\times S^{1}$ and assigning $\mathbb{Z}_{2}$ parity to objects in the theory is called the upstairs picture.  Since $S^{1}/\mathbb{Z}_{2}\simeq I$, where $I$ is the interval that can be coordinatized as $y\in[0,\pi R]$, we can work on the manifold with boundaries $\mathbb{R}^{4}\times I$, which is called the downstairs picture.  Boundary conditions on fields in this picture follow from parity assignments in the upstairs (though the mapping is not necessarily unique).  In this paper we'll work in the downstairs picture, based on upstairs picture parity assignments.

Fields and objects with $K$-indices can be assigned parities once the $\mathbb{Z}_{2}$ action is lifted from the base spacetime manifold to the $K$-bundle.  Fields with odd parity do not have independent zero modes in the 5D theory, nor independent modes on the 4D $\mathbb{Z}_{2}$ fixed-planes.  Splitting $i=(a,\alpha)$, the parity assignments for the bosonic fields are 
\begin{center}
\begin{tabular}{ccc}
Even & & Odd \\
$\hat{e}^{m}_{\mu}\;\;\hat{e}^{\bar{5}}_{5}$ & & $\hat{e}^{\bar{5}}_{\mu}\;\;\hat{e}^{m}_{5}$\\
$A^{\alpha}_{\mu}\;\;A^{a}_{5}\;\;A^{0}_{5}$ & & $A^{\alpha}_{5}\;\;A^{a}_{\mu}\;\;A^{0}_{\mu}$\\
$h^{0}\;\;h^{a}$ & & $h^{\alpha}$
\end{tabular}
\end{center}
In particular, the bare and physical 4D graviphoton, $A^{0}_{\mu}$ and $h_{I}A^{I}_{\mu}$, have odd parity. The gauge group on the boundaries and low energy effective theory is broken to a compact group $K_{\alpha}\subset K$ with gauge fields $A^{\alpha}_{\mu}$. 
Since $C_{IJK}$ is a rank-3 symmetric $K_{\alpha}$-invariant, we can write the $C_{ijk}$ in Eq.~(\ref{canonical}) such that
\begin{equation}
C_{000}=1,\;\;\;\;\;C_{00i}=0,\;\;\;\;\;C_{\tilde{I}\alpha\beta}=-\frac{1}{2}C_{\tilde{I}}\delta_{\alpha\beta},\;\;\;\;\;\theta(x^{5})C_{\alpha\beta\gamma},
\label{4dCtensor}\end{equation}
with the remaining components of $C_{ijk}$ unspecified.  We have split $a=(\hat{a},\tilde{a})$ such that $\hat{a}$ and $\tilde{I}=(0,\tilde{a})$ are non-singlet and singlet indices of $K_{\alpha}$, respectively.   In terms of these new indices, $i=(\alpha,\hat{a},\tilde{a})$. The $C_{\tilde{I}}$ are real-valued constants with $C_{0}=1$. The $\mathbb{Z}_{2}$-odd components have been redefined in terms of even ones via the $\mathbb{Z}_{2}$-odd distribution $\theta(x^{5})$, which is $-1$ for $x^{5}\in(-\pi R,0)$ and $+1$ for $x^{5}\in(0,\pi R)$. We can write the $K$-structure constant in a similar fashion.

\subsection{The approximate 4D effective theory}\label{sec:Leff}
From here on, we work in the downstairs picture.    
We take the f\"{u}nfbein to be parametrized as
\[
\hat{e}^{\hat{m}}_{\hat{\mu}}=
\left(
\begin{array}{ccc}
c^{\frac{1}{2}}e^{-\frac{\sigma}{2}} e^{m}_{\mu} &  & 2\hat{\kappa}e^{\sigma} C_{\mu}\\
0 & & e^{\sigma}
\end{array}
\right), 
\]
in which case the proper separation between boundaries is $r=\int e^{\vev{\sigma}}dy$.
We've included the free parameter $c$ of classical 4D Weyl transformations; $c=1$ is chosen in dimensional reductions, while $c=e^{\vev{\sigma}}$ is natural in compactifications~\cite{Nilles:1998uy}.  We expand $\sigma=\vev{\sigma}+\bar{\sigma}$, where $\bar{\sigma}$ is the ``fluctuation".  Furthermore, we'll parametrize the $y$-dependence of $e^{m}_{\mu}(x^{\mu},y)$ as $e^{-\gamma(y)}e^{m}_{\mu}(x^{\mu})$.  Overall, we will approximate by ignoring $y$-dependent fluctuations.  These choices put the action in the Einstein frame.

The effective 4D bosonic Lagrangian is obtained by imposing equations of motion for $\mathbb{Z}_{2}$-odd fields and integrating over $y$.  For our purposes it suffices to use the equations of motion for the $F^{I}_{\mu 5}$, truncating out the other odd fields.  Defining
\[\mathcal{I}:=\int^{\pi R}_{0} e^{\vev{\sigma}}e^{2\gamma} dy,\]
the equations of motion imply the replacements
\[
\mathcal{F}^{a}_{\mu 5}\rightarrow \frac{\pi R\, e^{\vev{\sigma}}e^{2\gamma}}{\mathcal{I}}D_{\mu}A^{a}\;\;\;\;\;\mbox{and}\;\;\;\;\;
\mathcal{F}^{\alpha}_{\mu 5}\rightarrow \frac{\pi R\, e^{\vev{\sigma}}e^{2\gamma}}{\mathcal{I}}\{A^{\alpha}_{\mu}(0)-A^{\alpha}_{\mu}(\pi R)\},
\]
where $D_{\mu}A^{a}=\partial_{\mu}A^{a}+A^{\alpha}_{\mu}f^{a}_{\alpha b}A^{b}$.
We've ignored $F\wedge F$ contributions, which would give higher derivative couplings in the effective Lagrangian.
In the first expression, the field $A^{a}$ on the right hand side can be expanded into a 4D vacuum expectation value and fluctuations $\vev{A^{a}}+\bar{A}^{a}$. It arises from the coordinate Wilson line phase $W^{a}=\int A^{a}_{5}\,dy=\pi R\,A^{a}$, taking a $y$-independent argument.  The second expression is not important in this paper; if the boundary conditions for the $A^{\alpha}_{\mu}$ are different at $y=0$ and $y=\pi R$, there will be a ``Scherk-Schwarz" mass term for the 4D gauge fields.   
 
Defining
\begin{equation}
\mathcal{A}:=C_{\tilde{I}}A^{\tilde{I}},\;\;\;\;\;\;\;\;\mathfrak{h}:=C_{\tilde{I}}h^{\tilde{I}},\;\;\;\;\;\;\;\;\mathcal{J}:=\int^{\pi R}_{0} e^{\vev{\sigma}}e^{-2\gamma}dy, \label{AhJ}
\end{equation}
the effective 4D Lagrangian in our approximation is then 
\begin{equation}\begin{split} 
&e^{-1}\int dy \mathcal{L}=-\frac{1}{2\kappa^{2}}\mathcal{R}-\frac{3}{4\kappa^{2}}\partial_{\mu}\sigma\partial^{\mu}\sigma-\frac{3\stackrel{\circ}{a}_{ab}}{4\kappa^{2}} D_{\mu}h^{a}D^{\mu}h^{b}-\frac{e^{\bar{\sigma}}\bar{\mathfrak{h}}}{4g^{2}} \mathcal{F}^{\alpha}_{\mu\nu}\mathcal{F}^{\alpha\,\mu\nu}\\
&-\frac{1}{2}\left\{\frac{\pi^{2} R^{2}}{g^{2}r\vev{\mathfrak{h}}}\frac{e^{-2\bar{\sigma}}}{ \mathcal{I}}\right\}\stackrel{\circ}{a}_{ab}D_{\mu}A^{a}D^{\mu}A^{b}+\frac{e^{-1}\kappa}{2\sqrt{6}}\frac{\pi R}{g^{3}}\frac{\mathcal{J}^{1/2}}{r^{3/2}\vev{\mathfrak{h}}^{3/2}} \epsilon^{\mu\nu\rho\sigma}\mathcal{A}\, \mathcal{F}^{\alpha}_{\mu\nu}\mathcal{F}^{\alpha}_{\rho\sigma}\\
&-\frac{3}{4\kappa^{2}}\frac{e^{-3\bar{\sigma}}}{\mathcal{J}}\stackrel{\circ}{a}_{IJ}(A^{c}f^{I}_{cd}h^{d})(A^{e}f^{J}_{ef}h^{f})+\{\mbox{Terms with}\;a,b,\ldots\rightarrow 0\}, \label{Leff}
\end{split}\end{equation}    
where the 4D tree-level couplings $\kappa$, $g$ come from
\begin{equation}
\hat{\kappa}^{2}=\kappa^{2} \mathcal{J}\;\;\;\;\;\;\;\;\;\;\hat{g}^{2}=g^{2}\vev{\mathfrak{h}}r.
\label{couplings}\end{equation}
In the scalar potential, only $f^{\hat{a}}_{\hat{b}\tilde{I}}$ and $f^{\alpha}_{\hat{a}\hat{b}}$ contribute, and $\stackrel{\circ}{a}_{\alpha a}=0$ on the boundaries. \\ 
\textit{Remarks}: \\
(i) While the positivity of the metric for the 5D kinetic terms is guaranteed by the positivity of the cubic polynomial $\mathcal{V}>0$,  here we additionally require $C_{\tilde{I}}h^{\tilde{I}}>0$.  This is an artifact of imposing the boundary conditions $h^{\alpha}=0$ in the characteristic polynomial $\mathcal{V}$.  When this is ultimately taken in the proper context of a larger supergravity theory (see~\cite{Mohaupt:2001be}), this condition may again follow from positivity of the new polynomial $\mathcal{V}'>0$.\\
(ii) The canonical 4D complex scalars are $z^{a}=A^{a}+i\kappa^{-1}\tilde{h}^{a}$, where $\tilde{h}^{a}=e^{\sigma}h^{a}$ (and similarly for $z^{0}$).  It's easier to do calculations before these definitions are imposed.  

\section{A QCD axion?}\label{sec:qcdaxion}

In the special class of $4D$ theories arising from five dimensions, axions come from 5D vectors in $\mathcal{F}\mathcal{F}A$ Chern-Simons type terms, which are present in classical supergravity.  Upon dimensional reduction, the generic ``axion/dilaton" $h_{I}z^{I}$ sits in a ``universal" 4D $\mathcal{N}=2$ vector multiplet, and parametrizes $SU(1,1)_{G}/U(1)$~\cite{mizo}.  The ``axion" in that case is the scalar $h_{I}A^{I}\equiv \Re (h_{I}z^{I})$ arising from the physical 5D graviphoton.  On $S^{1}/\mathbb{Z}_{2}$, the $K_{\alpha}$-singlet $\mathcal{A}$ appearing in Eq.~(\ref{Leff}) is the background-independent combination $C_{\tilde{I}}A^{\tilde{I}}$.  The superpartner saxion is $\tilde{\mathfrak{h}}:=e^{\sigma}C_{\tilde{I}}h^{\tilde{I}}$. The 5D action can be made invariant under local abelian transformations of the vectors~\cite{McReynolds:2007tt}, which act on the tower of Kaluza-Klein axions on the boundaries.  But the theory has neither a rigid nor local $U(1)_{PQ}$ symmetry associated with $C_{\tilde{I}}A^{\tilde{I}}$ at the zero mode level, and therefore also with the zero mode axion on the boundaries.

\subsection{The axion coupling strength}\label{sec:Mpq}

To estimate the effective axion couplings, we'll rescale the $K_{\alpha}$-singlet scalars to put their kinetic terms in canonical form.  First, we rescale 
\[
A^{\tilde{a}}\rightarrow \frac{\hat{g}\,\mathcal{I}^{1/2}}{\pi R}A^{\tilde{a}}.
\]
Assuming $\stackrel{\circ}{a}_{IJ}$ to be regular in the neighborhood of the fixed points, it takes the form
\begin{equation}
\stackrel{\circ}{a}_{IJ}=\{3C_{IKL}C_{JMN}h^{K}h^{L}h^{M}h^{N}-2C_{IJK}h^{K}\},
\label{aidentity}\end{equation}
where the $C_{IJK}$ are as in Eq.~(\ref{4dCtensor}).

 In the special class of theories with $C_{ijk}=0$ in Eq.~(\ref{canonical}), $C_{\tilde{a}}=0$ so that $\mathcal{A}=A^{0}$ and $\mathfrak{h}=h^{0}$.  From Eq.~(\ref{Leff}) it's clear that the final rescaling of $A^{0}, h^{0}$ to obtain canonical kinetic terms leaves the $\mathcal{A}FF$ term unchanged.  So in this restricted class of theories (and in the canonical basis), the third line of Eq.~(\ref{Leff}) becomes
\begin{equation}
\mathcal{L}_{\mathcal{A}}=\frac{1}{64\pi^{2}g^{2}}\left\{\frac{32\pi^{2}\kappa}{\sqrt{6}}\,\frac{(\mathcal{I}\mathcal{J})^{1/2}}{r\vev{h^{0}}}\right\}A^{0}\epsilon^{\mu\nu\rho\sigma}\mathcal{F}^{\alpha}_{\mu\nu}\mathcal{F}^{\alpha}_{\rho\sigma},
\label{Mpq}\end{equation}
where the quantity in brackets is $M_{PQ}^{-1}$.  In this basis, the effective coupling is $g^{2}\propto 1/h^{0}$, so we require $h^{0}>0$ (see remark (i) in Sect.~\ref{sec:Leff}).  From $\mathcal{V}=1$, $h^{0}$ never vanishes.  In fact, there are three branches where it can lie in the $h^{0}$-$h^{i}$ space, two of which have $h^{0}<0$, while the remaining has $h^{0}\geq 1$;  we therefore choose the positive branch.  We'll now specialize to two cases.\\
I. Flat bulk with $\gamma=0$:
\[
M_{PQ}=\frac{\sqrt{6}\kappa^{-1}}{32\pi^{2}}\,\vev{h^{0}}\geq \mathcal{O}(10^{16})\mbox{GeV}.
\]
This is a typical lower bound in higher dimensional scenarios, and as mentioned in the introduction, various assumptions can go into raising the allowed upper bound on $M_{PQ}$ to such a scale.  The 4D and 5D gravitational scales, $M_{4}:=\kappa^{-1}$ and $M_{5}:=\hat{\kappa}^{-2/3}$, and the proper separation $r$ between boundaries are (at tree level) related by $r[M_{5}]^{3}=[M_{4}]^{2}$.\\
II. Warped bulk with $\vev{\sigma}=\sigma_{0}$ constant and $\gamma=\Lambda z$.\footnote{This is not a ground state of the pure YMESGT, but can be obtained e.g. by gauging a subgroup of $SU(2)_{R}$.} \\
The line element is
\[
ds^{2}=e^{-2\Lambda z}\eta_{\mu\nu}dx^{\mu}dx^{\nu}+dz^{2},
\]
with the proper coordinate $z=e^{\sigma_{0}}y$ (such a background requires an extension of the pure YMESGT).  The axion scale is
\[
M_{PQ}=\frac{\sqrt{3}}{16\pi^{2}}\frac{\vev{h^{0}}\kappa^{-1}(r \Lambda)}{(\cosh[2 r\Lambda]-1)^{1/2}}.
\]
Taking $\vev{h^{0}}\sim\mathcal{O}(1)$, the window $10^{10}\mbox{GeV}\leq M_{PQ}\leq 10^{12}\mbox{GeV}$ for standard axion scenarios corresponds to $13\geq r\Lambda \geq 18$. The tree-level scales are related as $r[M_{5}]^{4}/(2\Lambda)\simeq [M_{4}]^{2}$.  

For theories with $C_{ijk}\neq 0$, $\mathcal{A}$ and $\mathfrak{h}$ take the general form in Eq.~(\ref{AhJ}). The canonical rescaling of the singlet scalars now changes the form of the $\mathcal{A}FF$ coupling, and generally the physical axion is split into several axions, which couple differently.  However, to rescale the individual scalars $A^{0}, A^{\tilde{a}}, h^{0}, h^{\tilde{a}}$, we need to turn to an approximation at lowest order in a $\hat{\kappa}$ expansion.  To do this, we express things in terms of special coordinates $\check{h}^{I}:=\tilde{h}^{I}/h^{0}$, $h^{0}\neq 0$, in which case the 5D scalars $\phi^{i}$ appear in $\check{h}^{i}=\hat{\kappa}\phi^{i}$ (and $\check{h}^{0}=1$). 
For $\vev{\phi^{\tilde{a}}}\ll \hat{\kappa}^{-1}$, the metric components of interest become
\[
\stackrel{\circ}{a}_{00}\simeq 3(h^{0})^{4}-2h^{0}\equiv H,\;\;\;\;\;\;
\stackrel{\circ}{a}_{\tilde{a}\tilde{b}}\simeq h^{0}\delta_{\tilde{a}\tilde{b}},\;\;\;\;\;\;\stackrel{\circ}{a}_{\tilde{a}0}\,\simeq 0.
\]  
In this ``quasi-rigid" limit, the theory lies in a neighborhood of the canonical basepoint of the 5D scalar manifold $h^{0}=1,h^{i}=0$. Then the fields are canonically rescaled as
\[\begin{split}
A^{0}&\rightarrow H^{-1/2}A^{0},\;\;\;\;\;\;\;\;\;\;h^{0}\rightarrow H^{-1/2}h^{0}\\
A^{\tilde{a}}&\rightarrow (h^{0})^{-1/2}A^{\tilde{a}},\;\;\;\;\;\;\;h^{\tilde{a}}\rightarrow (h^{0})^{-1/2}h^{\tilde{a}}. 
\end{split}\]   
Then Eq.~(\ref{Mpq}) holds with
\[
A^{0}\rightarrow A^{0}+[3(h^{0})^{2}-2]^{1/2}C_{\tilde{a}}A^{\tilde{a}}.
\]
\textit{Remark}: If bulk hyper- or vector multiplets are integrated out, the form of the $C_{IJK}$ of the effective 5D theory change; therefore the form of the effective axion coupling changes.  However, the strength of the coupling is unchanged.  This is also true in the reverse situation in which we integrate in such multiplets,for example, when the moduli space is singular for some points of the spacetime. 

\subsection{The scalar potential}\label{sec:scalarpot}

The axion is a linear combination of $K_{\alpha}$-singlet scalars $A^{\tilde{I}}$.  However, they generally appear in the non-negative potential coming from 5D $K$-coupling terms. In the 4D effective theory, there is the term\footnote{In the remainder of the paper, we'll normalize the 4D kinetic terms for $h^{a},A^{a}$ only up to the $\stackrel{\circ}{a}_{ab}$ non-linear $\sigma$-model metric; this is sufficient for our purposes.}
\[
e^{-1}\mathcal{L}_{eff}\sim -\frac{3g^{2}}{4\kappa^{2}}\frac{\mathcal{I}\mathcal{K}}{\mathcal{J}}\frac{\vev{\mathfrak{h}}r}{(\pi R)^{2}}\stackrel{\circ}{a}_{\hat{a}\hat{b}}e^{-3\bar{\sigma}}(A^{\tilde{I}}f^{\hat{a}}_{\tilde{I}\hat{c}}h^{\hat{c}})(A^{\tilde{J}}f^{\hat{b}}_{\tilde{J}\hat{d}}h^{\hat{d}}),
\]  
where the background satisfies $\stackrel{\circ}{a}_{\hat{a}\hat{b}}>0$ for all $\hat{a},\hat{b}$ and 
\[
\mathcal{K}:=\int e^{-4\gamma}e^{-\vev{\sigma}} dy.
\]
The coupling is of order $g^{2}$ for a flat bulk and $g^{2}e^{6\Lambda r}$ for a warped bulk.   

The bare 5D graviphoton $A^{0}_{\hat{\mu}}$ will not have gauge couplings when the 5D gauge group $K$ is compact since $f^{0}_{IJ}$ vanishes identically.  In that case, we don't have to worry about $A^{0}$ appearing in the potential.  Other contributions to $f^{\tilde{I}}_{\hat{a}\hat{b}}$ arise from additional would-be abelian factors in $K_{\alpha}$ that we break; e.g. if we break $K=SU(n+1)$ to $K_{\alpha}=SU(n)$, the scalars in the $\mathbf{n}\oplus\mathbf{\bar{n}}$ of $K_{\alpha}$ are charged with respect to the broken $U(1)$.  

If not all of the $f^{\tilde{I}}_{\hat{a}\hat{b}}$ vanish identically, $\vev{\mathcal{A}}$ is generally shifted from its QCD-instanton induced value, which would ruin the resolution to the strong-CP problem.  When $\vev{h^{\hat{a}}}=0$, the symmetries of the classical theory up to Peccei-Quinn shifts are $k_{\alpha}\oplus t^{\tilde{I}}$, where $k_{\alpha}$ is the 4D gauge algebra and $t^{\tilde{I}}$ are constant shifts of the $A^{\tilde{I}}$; the Peccei-Quinn shifts are due to the combination $C_{\tilde{I}}t^{\tilde{I}}$. However, further analysis of what the low energy potential will be requires non-perturbative contributions and a particular supersymmetry breaking scenario.  

\subsection{Bulk matter couplings}\label{sec:matter}

We have not specified where the Standard Model fermions are to come from.  Here, we'll consider matter coming from the bulk, in which case it sits in $n_{H}$ 5D hypermultiplets $\{\zeta^{A}, q^{X}\}$, where $X,A=1,\ldots,n_{H}$ (see~\cite{Ceresole:2000jd} for a recent discussion).  The scalars $q^{X}$ parametrize a quaternionic manifold $\mathcal{M}_{Q}$ such that the total scalar manifold of the 5D theory is $\mathcal{M}_{R}\times \mathcal{M}_{Q}$.  The gauge group $K$ must then be a common subgroup of the isometry group of each factor.  There are then terms in the Lagrangian     
\[
\hat{e}^{-1}\mathcal{L}_{5}\sim -\bar{\zeta}^{A}\Gamma_{\hat{\mu}}D_{\hat{\nu}}\zeta_{A}\,\hat{g}^{\hat{\mu}\hat{\nu}}+\frac{\sqrt{6}i}{8}\frac{\hat{\kappa}}{\hat{g}}h_{I}\bar{\zeta}_{A}\Gamma_{\hat{\mu}\hat{\nu}}\mathcal{F}^{I}_{\hat{\rho}\hat{\sigma}}\zeta^{A}\,\hat{g}^{\hat{\mu}\hat{\rho}}\hat{g}^{\hat{\nu}\hat{\sigma}},
\] 
where $h_{I}=C_{IJK}h^{J}h^{K}$.
In addition, there are scalar potential contributions, and the situation is similar to that of Sect.~\ref{sec:scalarpot}.
We are interested in the terms in the effective Lagrangian involving 4D matter fermions and their couplings to the scalar $\mathcal{A}$.  This involves a rescaling of $\zeta^{A}\rightarrow \mathcal{J}^{-1/2}\zeta^{A}$ to obtain a canonical kinetic term.  Then
\[
e^{-1}\mathcal{L}_{eff}\sim -e^{\bar{\sigma}}\bar{\zeta}^{A}\Gamma^{\mu}D_{\mu}\zeta_{A}-\frac{\sqrt{6}i}{8}\frac{\kappa\pi R}{ (\mathcal{I}\mathcal{J})^{1/2}}(h^{a}h^{a})e^{-3\bar{\sigma}}\,\partial_{\mu}\mathcal{A}\, \bar{\zeta}^{A}\Gamma^{\mu 5}\zeta_{A},
\] 
where the repeated index $a$ indicates a sum over it's values.  
The coefficient of the coupling term does not involve $M_{PQ}^{-1}$ as in the standard axion scenario. 

To determine whether this is significant, we should look at the standard calculations of axion mass and pion-axion coupling (for a textbook discussion, see~\cite{Weinberg:1996kr}).  First, the $\mathcal{A}\mathcal{F}\mathcal{F}$ term can be made to vanish by a local chiral transformation of the $\zeta^{A}$ (since this shifts the bare instanton vacuum parameter $\theta$).  This induces a change in the observable masses for the $\zeta^{A}$, as well as in the above derivative interaction.  In terms of ``up" $\zeta^{u}$ and ``down" $\zeta^{d}$ flavor quark fields in $\bar{\zeta}^{A}\Gamma^{\mu 5}C_{AB}\zeta^{B}= C_{u}\bar{\zeta}^{u}\Gamma^{\mu 5}\zeta^{u} + C_{d}\bar{\zeta}^{d}\Gamma^{\mu 5}\zeta^{d}+\cdots$, the original interaction can be written in the standard form as 
\begin{equation}
e^{-1}\mathcal{L}_{eff}\sim \,i\frac{f_{u}}{M_{PQ}}\partial_{\mu}\mathcal{A}\,\bar{\zeta}^{u}\Gamma^{5}\Gamma^{\mu}\zeta^{u}+i\frac{f_{d}}{M_{PQ}}\partial_{\mu}\mathcal{A}\,\bar{\zeta}^{d}\Gamma^{5}\Gamma^{\mu}\zeta^{d},
\label{barecoupling}\end{equation}
where the dimensionless parameters are
\begin{equation}
f_{u,d}=\frac{3}{128\pi^{2}}\frac{r\,\pi R}{\mathcal{I}\mathcal{J}}\vev{\mathfrak{h}}\left(h^{a}h^{a}\right) C_{u,d}.
\label{fvalue}\end{equation} 
Note that, since $h^{a}h^{a}\geq 0$, this coupling can vanish.  The chiral transformation shifts the coupling so that the final derivative interaction is of the form in Eq.~(\ref{barecoupling}) but with $f_{u,d}\rightarrow \tilde{f}_{u,d}\equiv f_{u,d}-c_{u,d}/2$, where $c_{u,d}$ are only restricted to satisfy $c_{u}+c_{d}=1$.  The effective axion-pion mixed kinetic term becomes
\[
e^{-1}\mathcal{L}_{eff}\sim -\frac{\tilde{f}_{u}-\tilde{f}_{d}}{M_{PQ}}F_{\pi}\;\partial_{\mu}\bar{\mathcal{A}}\,\partial^{\mu}\pi,
\]   
where $\pi$ is the pion field, $\mathcal{A}=\vev{\mathcal{A}}+\bar{\mathcal{A}}$, and $F_{\pi}=184\,\mbox{MeV}$ is the characteristic pion energy scale.  The $c_{u,d}$ can be chosen so that the axion and pion kinetic terms are diagonalized ($\tilde{f}_{u}=\tilde{f}_{d}$): 
\begin{equation}
c_{u}=1/2+f_{u}-f_{d},\;\;\;\;\;\;c_{d}=1/2+f_{d}-f_{u}.
\label{cvalues}\end{equation}
Since the standard calculations of pion and axion masses, as well as pion-axion interactions, involve the assumption that the $c_{u,d}$ are roughly $\mathcal{O}(1)$, we should check whether this holds in the present case in which the $f_{u,d}$ in Eqs.~(\ref{fvalue}) and~(\ref{cvalues}) are not $\mathcal{O}(1)$.  For flat bulk, 
\begin{equation}
f_{u,d}= \frac{3}{128\pi^{2}}\frac{\vev{\mathfrak{h}}(h^{a}h^{a})}{r M_{5}}C_{u,d}\,,
\label{fvalues}\end{equation}
where we're working in units in which $(\pi R)^{-1}\hspace{-0.5mm}=\hspace{-0.5mm} M_{5}$. For an effective 5D description to hold, we assume that $r M_{5}\geq \mathcal{O}(1)$; that is, the proper size of the fifth dimension is larger than the 5D fundamental (gravitational) distance scale.  As long as $\vev{\mathfrak{h}}(h^{a}h^{a})\leq \mathcal{O}(1)$, we have $f_{u,d}\leq \mathcal{O}(1)$ (this is certainly true in the quasi-rigid limit considered in Sect.~\ref{sec:Mpq}).  For a warped bulk as in Sect.~\ref{sec:Mpq}, Eq.~(\ref{fvalues}) is an upper bound. Therefore $c_{u,d}\sim \hspace{-0.5mm}1/2$ so that the standard estimates of the axion mass and axion-pion couplings for small $F_{\pi}/M_{PQ}$ holds.

\section{Conclusion}

We've considered classical 5D Yang-Mills Einstein supergravity on a spacetime with boundaries on which a restricted set of gauge fields propagate.   
The ``QCD-type" axion arises from a particular linear combination of those 5D vectors that are singlets under the broken 4D gauge group (includes the graviphoton).  At low energies, one can consider a quasi-rigid limit in which the target space background lies near the canonical ``basepoint" and the complications of the sigma-model geometry disappear (the estimates that follow, however, should be more robust). In a flat bulk, the coupling scale of the axion(s), $M_{PQ}\geq \mathcal{O}(10^{16})\mbox{GeV}$, is governed only by the 4D Planck scale.  In a warped bulk, $M_{PQ}$ can be lowered to lie within standard window $10^{10}-10^{12}\mbox{GeV}$ if the proper separation $r$ of the boundaries is $13-18$ times the radius of curvature $\Lambda^{-1}$.  For compact 5D gaugings without rank-reducing boundary conditions, the only other axion couplings are derivative.  Otherwise, the singlets appear in a 4D scalar potential, which may ruin the strong-CP resolution.  In the warped case, the coupling in this potential is large.  While classical and perturbative contributions to the potential can vanish, one must determine how non-perturbative contributions or supersymmetry breaking change this.  We also considered the case in which matter comes from bulk hypermultiplets, in which case the derivative couplings involve a scale different from the usual $M_{PQ}$. We checked that these couplings nevertheless still allow the standard approximations of the low energy effective axion-pion couplings and axion mass.  We have not explicitly discussed the conditions under which these spacetime backgrounds are admitted, and in the case of the warped background, it remains to fix the hierarchy between $r^{-1}$ and $\Lambda$ (see e.g.~\cite{radion}).    
\vspace{2mm}\\
\textbf{Acknowledgements}\\
Work supported
 by the European Commission RTN program ``Constituents, Fundamental Forces and Symmetries of the Universe" MRTN-CT-2004-005104 and by INFN, PRIN prot.2005024045-002.

  \end{document}